January 2, 2008

# Numerical modeling of the effect of carbon dioxide sequestration on the rate of pressure solution creep in limestone: preliminary results

**Simulation numérique de l'effet du stockage souterrain de dioxyde de carbone sur la déformation des calcaires par dissolution sous contrainte: résultats préliminaires**

François Renard [1, 2], Elisabeth Gundersen [2],
Roland Hellmann [1], Marielle Collombet [1], Yvi Le Guen [1,3]

[1] LGIT-CNRS-OSUG, Université Joseph Fourier, Grenoble, France

[2] Physics of Geological Processes, University of Oslo, Norway

[3] Institut Français du Pétrole, Rueil-Malmaison, France

F. Renard, R. Hellmann, M. Collombet & Y. Le Guen, LGIT, Université Joseph Fourier, BP 53X, 38041 Grenoble Cedex 9, France.

E. Gundersen & F. Renard, PGP, Institute of Physics, University of Oslo, postboks 1048, Blindern, 0316 Oslo, Norway.

Y. Le Guen, IFP, Division Géologie-Géochimie, 1-4 avenue de Bois Préau, 92500 Rueil-Malmaison, France.

e-mail: francois.renard@ujf-grenoble.fr, gundersen@geologi.uio.no, hellmann@obs.ujf-grenoble.fr, marielle.collombet@ujf-grenoble.fr, yleguen@lgit.obs.ujf-grenoble.fr.

**Abstract**

When carbon dioxide ($CO_2$) is injected into an aquifer or a depleted geological reservoir, its dissolution into solution results in acidification of the pore waters. As a consequence, the pore waters become more reactive, which leads to enhanced dissolution-precipitation processes and a modification of the mechanical and hydrological properties of the rock. This effect is especially important for limestones given that the solubility and reactivity of carbonates is strongly dependent on pH and the partial pressure of $CO_2$. The main mechanism that couples dissolution, precipitation and rock matrix deformation is commonly referred to as intergranular pressure solution creep (IPS) or pervasive pressure solution creep (PSC). This process involves dissolution at intergranular grain contacts subject to elevated stress, diffusion of dissolved material in an intergranular fluid, and precipitation in pore spaces subject to lower stress. This leads to an overall and pervasive reduction in porosity due to both grain indentation and precipitation in pore spaces. The percolation of $CO_2$-rich fluids may influence on-going compaction due to pressure solution and can therefore potentially affect the reservoir and its long-term $CO_2$ storage capacity. We aim at quantifying this effect by using a 2D numerical model to study the coupling between dissolution-precipitation processes, local mass transfer, and deformation of the rock over long time scales. We show that high partial pressures of dissolved $CO_2$ (up to 30 MPa) significantly increase the rates of compaction by a factor of 50 to 75, and also result in a concomitant decrease in the viscosity of the rock matrix.

**Résumé**

Lors de l'injection de dioxyde de carbone dans un réservoir déplété ou dans un aquifère, la dissolution du $CO_2$ dans l'eau de formation produit une acidification. Ce phénomène accélère les réactions de dissolution-précipitation avec la matrice rocheuse, et par conséquent, peut modifier notablement les propriétés mécaniques et hydrauliques des roches. De tels effets sont particulièrement importants dans les calcaires pour lesquels la solubilité et la réactivité des minéraux dépendent directement du pH, qui est lié à la pression partielle de $CO_2$. Le mécanisme de déformation par dissolution/précipitation sous contrainte est contrôlé par un couplage entre des processus de dissolution et de précipitation des minéraux et une déformation macroscopique de la matrice. Ce mécanisme implique une dissolution aux joints de grains où la contrainte normale est élevée, une diffusion de la matière dissoute dans le fluide intergranulaire, et une précipitation de matière dans les pores où la pression est plus faible. Cela induit une compaction de la roche et une diminution de porosité contrôlées à la fois par l'indentation des grains et par la précipitation dans les pores. La percolation de fluides riches en $CO_2$ tend à accélérer la compaction et peut ainsi modifier les propriétés mécaniques du réservoir à long terme. Dans cet article nous avons cherché à quantifier ce processus à l'aide d'un modèle numérique 2D qui couple les processus de dissolution et de précipitation à l'échelle des grains avec des transferts de matière à une échelle plus importante (quelques décimètres). Nous montrons que des pressions élevées de $CO_2$ (jusqu'à 30 MPa) accélèrent la vitesse de compaction des roches calcaires d'un facteur ~50 à ~75 et diminuent aussi leur viscosité.

# Introduction

The subsurface sequestration of $CO_2$ in geological repositories is frequently cited as a promising solution for reducing the amount of anthropogenically-produced $CO_2$ in the atmosphere. Some of the important issues involved in the long-term sequestration of $CO_2$ in such sites are discussed in an overview by Wawersik et al. (2001), for example. Therein it has been judged essential that models need to predict $CO_2$ sequestration behavior over time periods of several thousand years, which is the same order of magnitude as some climatic cycles. Therefore, in order to advance our knowledge of processes involved in the geological sequestration of $CO_2$, one of the most important objectives confronting geoscientists is understanding and quantifying all of the mechano-chemical processes, at both short and long time scales, that are relevant to $CO_2$ storage in geological formations.

When considering $CO_2 - H_2O$ injection into a geological repository, the following general chemical reactions (Stumm and Morgan, 1996) can be used to describe the subsequent water-rock interactions, based either on a mineralogy dominated by aluminosilicates (Eq. 1) or calcium carbonate

$$\text{cation} - \text{Al} - \text{silicate} + CO_2 + H_2O \Leftrightarrow \text{cation} + HCO_3^- + H_4SiO_4 + \text{Al} - \text{silicate} \qquad (1)$$

$$CaCO_3 + CO_2 + H_2O \Leftrightarrow 2HCO_3^- + Ca^{2+} \qquad (2)$$

Since we examine $CO_2$ sequestration within the context of pervasive pressure solution creep (PSC) in limestone, Eq. (2) is thus of particular relevance to this study.

PSC is a mechano-chemical process characterized by ductile deformation and local mass transfer affecting water saturated porous rocks (e.g. Weyl, 1959; Rutter, 1976; Gratier and Guiguet, 1986; Dewers and Ortoleva, 1990; Spiers and Brzesowsky, 1993; Gundersen et al., 2002; Yasuhara et al., 2003). This ductile deformation mechanism occurs in the upper crust and plays an important role in the compaction of sedimentary rocks during diagenesis (Ortoleva, 1994; Tuncay et al., 2000). PSC is driven by differences in chemical potential induced by differential stress along grain surfaces in the rock matrix. PSC can be modeled as a serial process involving four successive steps:

- stress-enhanced dissolution at grain-grain interfaces subject to elevated normal stress;

- diffusion of dissolved material (solutes) through intergranular fluid films;

- precipitation of dissolved material in adjacent pore spaces (grain surfaces in contact with pore fluid);

- transport of dissolved material to distant pores, which can induce local mass transfer (Gundersen et al., 2002).

Since it is assumed that PSC operates as a serial process, the slowest step imposes the overall rate for deformation (Rutter, 1976; Gratz, 1991). The first three individual steps of PSC (dissolution, diffusion, precipitation) are in turn influenced by local parameters such as temperature, stress state of the rock matrix, fluid pressure, and fluid chemistry (Rutter, 1976). The PSC mechanism of deformation is slow and operates over long geological time scales. Because of this, it is even possible that the slow step can change from one process to another over time during compaction of the rock matrix.

Because PSC operates over geological time scales, correctly predicting the long-term stability of a $CO_2$ repository requires accurate modeling. PSC models are based for the most part on kinetic and equilibrium parameters derived from laboratory dissolution and precipitation experiments, as well as from pressure solution experiments that typically run for only a small fraction of the time scales associated with natural PSC deformation.

Injection of $CO_2$ causes chemical and flow regime perturbations that affect the PSC process, causing the porosity, permeability, and mechanical stability of the porous rock matrix to evolve over time. There are multiple reasons that are responsible for this. The acidification of pore fluids due to the dissolution of $CO_2$ generally increases rates of fluid-rock interactions. This is particularly important in limestones where a decrease in pH increases both the rate of calcite dissolution and calcite solubility. In addition, higher concentrations of dissolved calcium carbonate can result in increased rates of precipitation. The porosity of the rock matrix is reduced during PSC, due primarily to grain indentation and precipitation in pore spaces. Taken together, the dissolution, diffusion, and precipitation processes have the potential for modifying the long-term porosity and permeability of the repository rock, as well as its mechanical stability.

The deformation of chalk, both dry and in the presence of fluids, has been widely investigated experimentally. Most of the published work is based on chalk deformation data that have been interpreted solely in terms of mechanical processes (e.g., Botter, 1985; Da Silva et al., 1985; Jones and Leddra, 1989; Monjoie et al., 1990; Shao et al., 1994; Piau and Maury, 1995; Schroeder and Shao, 1996; Homand et al., 1998; Risnes and Flaageng, 1999). Recently, however, a few studies have examined chalk-fluid deformation within the context of chemical processes associated with PSC (Hellmann et al., 2002a, b; Heggheim et al., in press). In addition, PSC rates have also been measured for calcite aggregates (Zhang et al., 2002) and single calcite grains (Zubtsov et al., 2004).

In this study, we examine the effect of elevated concentrations of dissolved $CO_2$ on the overall PSC rate of limestone dominated aquifers or reservoirs at burial depths relevant to $CO_2$ storage (1000-3000 m). The model treats the post-injection phase of sequestration, where the aqueous pore fluids have been homogeneously acidified by the presence of $CO_2$. The partial pressure of $CO_2$ is fixed and remains constant for each simulation. The model also makes the approximation that the $p_{CO_2}$ equals the pore pressure in the reservoir. Using a 2D numerical model, we examine how various parameters such as grain size, burial depth, rock texture, and the partial pressure of $CO_2$ modify the rate of matrix compaction by pervasive PSC. Only the effect of $CO_2$ dissolved in water is considered (*i.e.* a single subcritical aqueous phase); we do not consider the injection of a supercritical $CO_2$ phase since the reactivity (i.e. solubility) of calcite therein is predicted to be minimal.

Below, we first present a brief review of the relevant thermodynamics and kinetics of the calcite-$H_2O$-$CO_2$ system that is used to model the dissolution-precipitation processes. This is followed by a description of the basics of our PSC model. Lastly, we present several results from the 2D simulations.

## 1. Calcite-$H_2O$-$CO_2$ thermodynamics and kinetics

### 1.1. *Conventions*

Before we address the thermodynamics of the calcite-$H_2O$-$CO_2$ system, it is important to note the conventions we use. First, even though equilibrium constants are by definition based on activities of product and reactant species and phases, we make a simplifying assumption of equating activities of aqueous species with concentrations (in mol m$^{-3}$); that is we assume that their activity coefficients are equal to 1. This approximation is based on the relatively low ionic strength of the solutions (the maximum ionic strength ($I$) = 0.115 molal, calculated using EQ3NR, Wolery, 1992, and the activity coefficient is greater than 0.85, see Fig. 1 of Kervevan *et al.*, 2005). Solid phases are always assigned an activity equal to 1. We also assume that the molarities are equal to the molalities and the density correction is neglected. Lastly, the use of the term 'fluid' has the same meaning as 'solution', no supercriticality is implied.

*1.2.    Overview of equilibrium thermodynamic relations at 25 °C and 0.1MPa*

In this study we consider the system $CaCO_3$-$H_2O$-$CO_2$ with the following equilibrium constants (Nordstrom et al., 1990) at standard temperature and pressure (STP = 25 °C and 0.1 MPa pressure; see Table 1 for symbols)

$$K_w = [H^+][OH^-] \qquad \log K_{eq}=-14.00 \qquad (3a)$$

$$K_{CO_2} = [H_2CO_3^*]/p_{CO_2} \qquad \log K_{eq}=-1.47 \qquad (3b)$$

$$K_1 = [H^+][HCO_3^-]/[H_2CO_3^*] \qquad \log K_{eq}=-6.35 \qquad (3c)$$

$$K_2 = [H^+][CO_3^{2-}]/[HCO_3^-] \qquad \log K_{eq}=-10.33 \qquad (3d)$$

$$K_{CaHCO_3^+} = [Ca^{2+}][HCO_3^-]/[CaHCO_3^+] \qquad \log K_{eq}=-1.11 \qquad (3e)$$

$$K_{sp} = [Ca^{2+}][CO_3^{2-}] \qquad \log K_{eq}=-8.48 \qquad (3f)$$

In Eqs. 3b,c $[H_2CO_3^*]$ denotes $[CO_2(aq)+[H_2CO_3]]$.

The following derived equilibrium constant expressions are also applied:

$$K_1 K_2 K_{CO_2} = [H^+]^2[CO_3^{2-}]/p_{CO_2} \qquad \log K_{eq}=-18.15 \qquad (3g)$$

$$K_{diss.,calcite} = [Ca^{2+}][HCO_3^-]^2/p_{CO_2} \qquad \log K_{eq}=-5.97 \qquad (3h)$$

Expression (3g) represents the dissolution/equilibration of $CO_2$ in $H_2O$, including dissociation reactions, while (3h) represents equilibrium of the $H_2O$-$CO_2$ system with respect to calcite, as shown in Eq. (2). Other aqueous species and reactions in this system, for example $CaOH^+$ and $CaCO_3^0$ and their respective dissociation reactions, are not considered here since their concentrations are insignificant at $p_{CO_2} > 10^{-4}$ MPa (see Fig. 6.5, Langmuir, 1997). The only exception to this is the species $CO_3^{2-}$ whose concentration, albeit low, is necessary for the calculation of the equilibrium constant $K_{sp}$.

In addition to the above equilibria, the following aqueous charge balance holds

$$2[Ca^{2+}]+[CaHCO_3^+]+[H^+]=[HCO_3^-]+2[CO_3^{2-}]+[OH^-] \qquad (3i)$$

At circum-neutral to acid pH conditions and $p_{CO_2} > 10^{-4}$ MPa, the species $CaOH^+$, $CO_3^{2-}$, and $OH^-$ are of minor importance and can be neglected in this charge balance.

Using the chemical equilibria and charge balance relations given in Eqs. 3a-i, the concentrations of all chemical species pertinent to the calcite-$H_2O$-$CO_2$ system at STP can be calculated. However, in order to be useful for PSC, the code recalculates these equilibrium concentrations to conform to the relevant temperature and pressure conditions of the fluids present either in the contact zone or in the pore space (as discussed in the following sections). Note, however, that the charge balance relation in Eq. 3i is not pressure or temperature dependent.

In order to simplify the numerical code, only the $Ca^{2+}$ concentration is allowed to vary as a function of time. All of the other species' concentrations remain fixed at their initial equilibrium values calculated for $t = 0$. Thus, at this stage in the development of the code, we have made the simplifying assumption that PSC depends only on one chemical species, such that only $Ca^{2+}$ dissolution, precipitation, and local transport are rate-determining.

*1.3.  Equilibrium thermodynamic relations at elevated temperatures and pressures*

1.3.1.  <u>Dependence on temperature</u>

An empirical expression for the solubility of calcite (Eq. 3f) as a function of temperature $T$ and salinity $S$ has been derived by Mucci (1983) and is given by

$$\log K_{sp} = -171.9065 - 0.077993\, T + \frac{2839.319}{T} + 71.595 \log T + \left(-0.77712 + 0.0028426\, T + \frac{178.34}{T}\right) S^{0.5} - 0.07711\, S + 0.0041249\, S^{1.5} \quad (4)$$

where $T$ is in Kelvin. As defined by Mucci (1983), this equation is valid for 0-40°C and $S$ = 5-35 g/kg. For the purposes of this study, we restrict ourselves to the first four terms of Eq. 4, which is the expression originally determined by Plummer and Busenberg (1982) and is applicable at 0-90°C. Using the abbreviated form of Eq. (4), log $K_{sp}$ at 25°C and 100°C ($P$ = 0.1 MPa) is equal to -8.479 and -9.264, respectively. The retrograde solubility of calcite is a consequence of the overall dissolution reaction having a negative enthalpy.

The general reactions describing $CO_2$ dissolution and dissociation of carbonic acid species in solution have a weak temperature dependence, due to small enthalpies of reaction. The temperature dependence of $K_{CO_2}$, $K_1$, $K_2$ and $K_{CaHCO_3^+}$ (Eqs. 3b, 3c, 3d, 3e, respectively) from 0-90 °C can be represented by an empirical relationship (Table A1.1, Langmuir, 1997; original reference: Plummer and Busenberg, 1982) of the same form as Eq. (4), that is

$$\log K_i = a + bT + \frac{c}{T} + d \log_{10} T + \frac{e}{T^2}. \qquad (5)$$

The values of *a*, *b*, *c*, *d*, and *e* in Eq. (5) are given in Table 2.

### 1.3.2. Dependence on pressure of the equilibrium constant of calcite dissociation

The expression for the pressure dependence of $K_{sp}$ (Eq. 3f) that we use is the following (Lown et al., 1968; Millero, 1982)

$$\ln(K_{sp}^P / K_{sp}^{P_0}) = -(\Delta V_r^0 / RT)(P - P_0) + (0.5 \Delta \kappa_r / RT)(P - P_0)^2. \qquad (6)$$

Here, $\Delta V_r^0$ and $\Delta \kappa_r$ are the molal volume and compressibility changes for the dissolution reaction, $P$ is the applied pressure in MPa, and $R = 8.32$ cm$^3$ MPa mol$^{-1}$ K$^{-1}$. Thus, using $\Delta V_{Ksp}^0 = -58.3$ cm$^3$ mol$^{-1}$ at 25°C (Langmuir, 1997) and $\Delta \kappa_r = -1.5 \cdot 10^{-3}$ cm$^3$ MPa$^{-1}$ mol$^{-1}$ at 25°C for the solubility of calcite in pure water (Owen and Brinkley, 1941), we can then calculate $K_{sp}$ at elevated pressures. As an example, a change in system pressure from $P_0$=0.1 to $P$=10 MPa results in the following increase in the solubility product: $K_{sp}^P / K_{sp}^{P_0} = 1.26$ at 25°C in pure water.

In the present version of the code, the pressure dependencies of the other equilibrium constants (Eqs. 3a-e, g-h) have not been considered.

### 1.4.    Overview of kinetic rate laws for calcite dissolution

The rates of calcite dissolution and precipitation in aqueous solutions have been actively investigated over many decades (*e.g.* see a thorough review by Morse and Arvidson, 2002; and references therein). One of the most widely used empirically-derived kinetic rate laws is based on the work of Busenberg, Plummer, and co-workers

(Plummer et al., 1978; Busenberg and Plummer, 1986), in which the rate of reaction (dissolution or precipitation) of calcite can be expressed as a function of three parallel forward rates and one backward rate

$$R = k_1[H^+] + k_2[H_2CO_3^0] + k_3[H_2O] - k_4[Ca^{2+}][HCO_3^-] \qquad (7)$$

where $R$ is the overall reaction rate, $k_1$, $k_2$, $k_3$, $k_4$ are kinetic rate constants, and $[i]$ represents the bulk solution concentrations (in mol/m$^3$ of water) of aqueous species $i$. When considering only the three forward reactions ($k_1$, $k_2$, $k_3$) in Eq. (7), the first kinetic term dominates at acid pH, the second term dominates at elevated $p_{CO_2}$ (> 0.01 MPa), while the third term dominates at pH > 6 and low $p_{CO_2}$ (< 0.01 MPa). In $p_{CO_2}$-pH space there is also a region where all three terms must be considered (see Fig. 12, Plummer et al., 1978).

In circum-neutral pH solutions, and close to calcite equilibrium ($\Omega > 0.6$), the dissolution reaction can be represented by a simpler alternative kinetic rate law for dissolution that has the following form (Berner and Morse, 1974; Busenberg and Plummer, 1986; Wollast, 1990; Hales and Emerson, 1997)

$$R = k_{diss}(1-\Omega)^n = k_{diss}(1-Q/K_{sp})^n. \qquad (8)$$

Here $k_{diss}$ is the rate constant for calcite dissolution (denoted by $k_5$ in Busenberg and Plummer, 1986), $(1-\Omega)$ is the chemical reaction affinity term (*i.e.* degree of solution undersaturation or oversaturation), $n$ is the reaction order, and $\Omega$ is defined as the ratio of the ion activity product $Q_p = a_{Ca^{2+}} \times a_{CO_3^{2-}} = [Ca^{2+}] \cdot [CO_3^{2-}]$ (see Section 1.1) to the equilibrium constant $K_{sp}$ (Eq. 3f). For calcite, $n = 1$ and log $k_{diss}$ = -9.93 at 25 °C and 0.1 MPa (Plummer et al., 1978, units of $k_{diss}$ in mol cm$^{-2}$ s$^{-1}$). A linear relation between the rate of dissolution and chemical affinity (i.e. $n \approx 1$) at conditions close to equilibrium has been reported in other studies, as well (e.g. Cubillas et al., 2004). The rate law given in Eq. (8) has in fact been described as, "the most commonly used equation in geosciences to describe the rate of carbonate mineral dissolution" (Morse and Arvidson, 2002). The kinetic expression in Eq. (8), which describes the net rate of reaction (dissolution and precipitation), has both a mechanistic and empirical origin (Wollast, 1990), based on the simple reaction $CaCO_3 \leftrightarrow Ca^{2+} + CO_3^{2-}$.

However, the simple, quasi-linear relationship between rate and chemical affinity shown in Eq. 8 is perhaps still debatable. As an example, Svensson and Dreybrodt (1992) report that for natural (i.e. impure) calcite, the dissolution rate-affinity relation deviates from a linear relation chose to equilibrium, yielding a value of $n \approx$ 3-4 in Eq. 8. In the following treatment, however, we neglect the uncertainty associated with this potential non-linearity, as it is much smaller than the uncertainty associated with the measurement of the value of $k_{diss}$ that we adopt.

*1.5.     Effect of temperature and pressure on the rate constant $k_{diss}$*

Elevated $T$ and $P$ conditions have two effects on the overall rate of calcite dissolution: they modify the rate constant $k_{diss}$, and also the chemical affinity term $(1 - \Omega)$. The dependence of $k_{diss}$ on $T$ can be expressed using the classical Arrhenius relation

$$k_{diss} = A \exp\left(\frac{-E_a}{RT}\right) \qquad (9)$$

where $A$ is a pre-exponential frequency factor, and $E_a$ is the overall thermal activation energy. Plummer et al. (1978) determined an $E_a$ = 33.05 kJ mol$^{-1}$ for the rate constant $k_3$ (Eq. 7); this corresponds to $A$ = 11.7. An $E_a$ value of 35 kJ mol$^{-1}$ was also reported by Sjöberg (1978). Morse and Arvidson (2002) proposed that activation energy values in this range are consistent with surface reaction control kinetics, but the value of 35 kJ/mol suggests a mixed control: activation energies in the range 0-25 kJ/mol are typical for diffusion in a fluid media, whereas 50 kJ/mol and greater is typical for surface reaction control (Lasaga, 1998). In the study of Busenberg and Plummer (1986), the values of $k_3$ (Eq. 7 above) and $k_{diss}$ (Eq. 8 above, equiv. to $k_5$) are very close in value, and therefore we base the temperature dependence of $k_{diss}$ on the following empirical expression for $k_3$ (25-48°C) given in Plummer et al. (1978)

$$log(k_3) = -1.10 - 1737/T \qquad (10)$$

The effect of fluid pressure (neglecting the effect of $p_{CO_2}$ -see next section) on the rate constant $k_{diss}$ is not considered, given the lack of pertinent data in the literature.

*1.6.     Effect of $p_{CO_2}$ on the calcite dissolution rate constant*

The effect of $p_{CO_2}$ on the dissolution rate constant $k_{diss}$, independent of the pH effect, is controversial. Several studies have reported a positive relation, *i.e.* a first order dependence between calcite dissolution rates and $p_{CO_2}$ at fixed pH and variable ionic strength (Plummer et al., 1978; Busenberg and Plummer, 1986; Arakaki and Mucci, 1995). However, Pokvrosky et al. (2004) have shown that increasing $p_{CO_2}$ up to 5 MPa results in just a modest increase in the calcite dissolution rate. They report that the kinetic rate constant increases by a factor of three as $p_{CO_2}$ is increased from 0.1 to 2 MPa, and in the range of $p_{CO_2}$ = 2-5 MPa, they measure no change in the dissolution rate constant. In addition, these authors have also shown that the ionic strength has no effect on the dissolution rate constant. In our model, we use the results of Pokrovsky et al. (2004) for defining the change in $k_{diss}$ as a function of $p_{CO_2}$.

## 2. The mathematical model of PSC

### 2.1. *Coupling between dissolution-diffusion-precipitation processes and matrix deformation*

Using our model we compute how the petrophysical properties of the rock matrix (permeability, porosity) and the volumetric strain (compaction) evolve with time due to PSC. The rock aggregate is a monomineralic limestone (i.e., pure calcite), with either a homogeneous or variable grain size (i.e. random spatial variation based on a uniform grain size distribution). The model treats the coupling between chemical reactions of calcite dissolution and precipitation, diffusion, and PSC deformation in two steps:

- textural equations at the grain scale allow for grain deformation due to chemical reactions occurring at contact and pore surfaces (including diffusion in the contact fluid);
- a mass conservation relationship at a global scale which takes into account the transfer of dissolved material by diffusion in the interconnected pore spaces.

Therefore, microscopic grain scale processes and macroscopic mass transport within the model volume are fully coupled (for a complete review, see Gundersen et al., 2002).

The basis of our model for treating the effect of elevated normal stress on individual grains comprising a rock matrix is the thermodynamic relationship which relates normal stress to chemical potential (Gibbs, 1878; Kamb, 1961; Paterson, 1973; Lehner, 1995; Renard et al., 1999). The higher chemical potential of calcite surfaces in the contact zone results in a higher solubility with respect to calcite surfaces of the pore space. Ultimately, this difference in chemical potentials drives two processes: stress-enhanced dissolution of the contact zone surfaces and diffusion of dissolved material from the contact zone out to the pore fluid. The diffusion of dissolved material occurs within a trapped water film separating the grains (Rutter, 1976; de Meer et al., 2002; Dysthe et al., 2002b).

Finally, grain-scale dissolution-precipitation processes in pore spaces are coupled to bulk diffusion of solute within the entire porous medium. The conservation relationships are derived by a global mass balance of the solute phase in the pore volume and a local mass balance at each grain contact (Gundersen et al., 2002). These relationships are then coupled to equations which express the deformation of the grains and the evolution of the rock texture.

2.2.     *Deformation of individual grains and rock matrix*

In our model the rock matrix is a 20x20 cm domain, located at a specified depth. The geological conditions (stress at the boundaries, pore pressure, temperature) are assumed to stay constant and are chosen to represent relevant conditions for $CO_2$ sequestration (depths between 1 and 3 km, see Table 3).

The rock matrix is modeled as a network of solid calcite grains with well-defined grain-grain contacts and interconnected pore spaces between grains (Fig. 1). The aggregate grain framework is represented by a dense cubic packing of truncated spheres that represent the individual grains. In Fig. 1 it is also important to distinguish the two different types of grain-fluid interfaces that are treated in the model:
- grain surfaces at intergranular (i.e. grain-grain) contacts that are separated by a trapped fluid film,
- 'free' grain surfaces that are in contact with pore fluids.

The intergranular grain boundary is considered as a flat interface (Hickman and Evans, 1995) with a mean roughness of several nanometers averaged over the contact.

This interfacial contact zone contains a trapped fluid film that is postulated to have the unusual characteristic of being able to support a shear stress transmitted by the normal stress imposed on the grains. In addition to nanoscale topography, there is also ample evidence that grain-grain contact zones have a structure of channels and islands at a micrometer scale (Hickman and Evans, 1992; Schutjens and Spiers, 1999; Renard et al., 1999; Dysthe et al., 2002a). The exact relation between these nano- and microscale features is not yet well understood.

The initial radius of the spherical grains, $L_f$, can vary between the different elements. The other lengths, describing the truncations of the spheres, $L_x$ and $L_z$, are given as $2 \cdot 0.9 \cdot L_f$. The choice of this value allows initial porosities close to 30%. Each simulation lasts until the porosity in the whole simulation domain is less than a threshold value equal to 5%. Below this value, transport by diffusion between the pores ceases and pressure solution only continues locally inside this element, as a closed system. The decrease in porosity from an initial 30% to a final value of 5% represents a finite volumetric strain of approximately 20%. The time associated with this porosity reduction defines the average rate of deformation for PSC and forms the basis for comparing the various cases examined (variable depth, $p_{CO_2}$)

The entire matrix of grains with cubic packing is subjected to a constant normal stress component ($\sigma_n$). We then define a contact normal stress component ($\sigma_c$) as the mean stress at each grain contact surface. This stress depends on the relationship between the surface area and the diameter of the sphere. This relationship is independent of the initial size of the individual grains (Dewers and Ortoleva, 1990), and the stress at the contact is proportional to some (positive) power of the porosity.

The evolution of the rock texture, which leads to a compaction of the aggregate, is computed as a result of the coupling between dissolution, diffusion, precipitation and mass transport in the fluid. This textural model allows us to take into account the sequential coupling between chemical processes (dissolution at grain contacts, diffusion, and precipitation on free pore surfaces) and the mechanical evolution (deformation of the grains) of the rock matrix. The resulting model is then a set of highly coupled non-linear equations that can only be solved using numerical methods. All of the parameters used in the following equations and their units are given in Table 1.

*2.3. Matrix-fluid chemical equilibrium at t = 0*

The initial state of the model is based on two separate equilibrium chemical states, such that at $t = 0$, the rock matrix is in equilibrium with, respectively, the contact and pore fluids. The initial fluid compositions are determined as a function of the specified temperature, pressure (i.e. depth), and $p_{CO_2}$ (see sections 1.1 and 1.2). Even though $T$ and $p_{CO_2}$ are the same in the contact zone and pore space for any specified depth, the respective fluid pressures will differ since the fluid film in the contact zone supports the intergranular normal stress, which is higher than the pore fluid pressure (i.e. $P_c > P_p$). Thus, the code starts with two initial local equilibrium conditions: one between the calcite in the contact zone and the fluid film, the other between the calcite in the pore space and the pore fluid. The initial concentrations of the five most important aqueous species (see sections 1.2., 1.3), $Ca^{2+}$, $CaHCO_3^+$, $HCO_3^-$, $CO_3^{2-}$, and $H^+$, are calculated for both the contact zone fluid and the pore space fluid. Given that $P_c > P_p$, the concentrations of these five species will not be the same in the contact zone fluid and in the pore fluid. Thus, an initial chemical imbalance exists between the contact zones and the pore spaces.

The evolution of the chemistry of the contact fluid and the pore fluid is treated in a different manner once deformation starts, that is for $t > 0$. The initial chemical imbalance between the contact zone fluid and the pore space fluid drives diffusion that initiates the PSC deformation of the rock matrix. This diffusion process, which is constrained by the code to depend only on the negative $[Ca^{2+}]$ gradient between the contact zone and the pore space ($[Ca^{2+}]_c > [Ca^{2+}]_p$), occurs within the intergranular fluid film in the contact zone. Once initiated by diffusion in the contact zone, PSC deformation of the rock matrix continues to evolve with time. In doing so, the code has no *a priori* constraints with respect to whether dissolution, diffusion, or precipitation is rate determining.

The code is presently configured such that once deformation starts ($t > 0$), only the $[Ca^{2+}]$ is allowed to change throughout the rock matrix. The concentrations of the other chemical species do not change as a function of time; they retain their initial, respective equilibrium values in the contact zone fluid and pore space fluid throughout the deformation process. Therefore, PSC depends on only one chemical species, and as a consequence, $Ca^{2+}$ dissolution, diffusion, or precipitation is rate-determining. This simplication in the code was necessary to ensure sufficient numerical stability of the results. Since calcite dissolution and precipitation, as shown in section 1.4, also depends on $[CO_3^{2-}]$, the code uses its fixed, initial value where appropriate.

The pore fluid is considered to be an 'open' system, permitting the exchange of mass within the porous medium. Most importantly, this implies that the $p_{CO_2}$ remains constant with respect to time and the matrix position and does not depend on the reaction progress of $CaCO_3$ dissolution in the contact zone or precipitation on pore surfaces.

*2.4. Conservation equations in the pore fluid*

Defining $c_{p,Ca}$ to represent the concentration of $Ca^{2+}$ in the pore fluid, then $\phi \cdot c_{p,Ca}$ (where $\phi$ is the porosity) is the concentration of $Ca^{2+}$ in the total model volume. The rate of change, given by $\phi \dfrac{\partial c_{p,Ca}}{\partial t}$, is the net result of:

- diffusion in the pore fluid;

- addition or loss of $Ca^{2+}$ into the fluid by, respectively, stress-enhanced dissolution at the grain contacts or removal by precipitation in the pore spaces.

The mass balance for the $Ca^{2+}$ phase in the pore volume is then

$$\phi \frac{\partial c_{p,Ca}}{\partial t} = D_p \nabla^2 c_{p,Ca} + \frac{1}{L_x L_y L_z}\left[\frac{\partial m_{prec,Ca}}{\partial t} + 2\sum_{i=1}^{3}\frac{\partial m_{diff,i,Ca}}{\partial t}\right] \qquad (11)$$

where the lengths $L_x$, $L_y$, $L_z$ are geometric grain parameters defined in Fig. 1, $D_p$ is the coefficient of diffusion in the pore fluid; $m_{prec,Ca}$ and $m_{diff,Ca}$ are the moles of precipitated $Ca^{2+}$ on the pore walls (i.e. pore free faces) and the solute moles derived from diffusion from the grain contacts to the pore fluid, respectively (see Table 1 for symbols). The first term on the right side in Eq. (11) describes the diffusive transport of solutes in the pore space by a standard Fickian diffusion relation.

The second term on the right hand side in Eq. (11) represents mass loss due to precipitation on the grain surfaces and is given as:

$$\frac{\partial m_{prec,Ca}}{\partial t} = k_{prec} A_p \left(1 - \frac{c_{p,Ca} c_{p,CO3}}{K_{sp,p}}\right) \quad (12)$$

where $k_p$ is the rate constant for calcite precipitation, $A_p$ is the grain surface area in contact with the pore fluid (reactive area), $c_{p,Ca} c_{p,CO3}$ is the ion activity product in the pore fluid, and $K_{sp,p}$ is the calcite solubility product in the pore fluid. Notice that as the geometry of the grains changes during compaction, $A_p$ varies as well. At conditions close to equilibrium, the rate expression used in the model to calculate the overall reactivity of calcite does not differentiate between dissolution and precipitation (i.e. $k_{diss} = k_{prec}$). This appears to be a reasonable approximation based on available experimental evidence at conditions very close to equilibrium (e.g. Fig. 8, Busenberg and Plummer, 1986).

*2.5. Conservation equations at the grain contacts and grain shape evolution*

The last term in Eq. (11) describes mass production due to diffusional mass transport out of the intergranular contact zone and reads:

$$\frac{\partial m_{diff,i,Ca}}{\partial t} = 2\pi \frac{\Delta}{2} D_c \left(c_{c,i,Ca} - c_{p,Ca}\right); \; i=x,y,z \quad (13)$$

where $c_{c,i,Ca}$ is the concentration of $Ca^{2+}$ in the grain contact perpendicular to axis $i$ ($i=x$, $y$, $z$). The driving force for this transport is a concentration gradient related to the difference in stress concentration at intergranular and free grain contacts. In the model, the length of the intergranular diffusion path relates directly to the rate of compaction (*i.e.* smaller grain matrices compact faster than larger grains). The transfer of dissolved $Ca^{2+}$ within the interfacial fluid from grain-grain contacts to the pore fluid occurs via a diffusion process. Thus, the chemistry of the interfacial fluid evolves not only by the addition of material via dissolution, but also by the removal of dissolved material by diffusion.

In general, the rate of ion diffusion ($D_c$) in thin, confined interfacial fluid films is not directly measurable, but recent studies (*e.g.* Dysthe et al., 2002b; Alcantar et al., 2003) conclude that the rate is no more than two orders of magnitude slower than the corresponding bulk fluid value (at a given $T$ and $P$). The adopted value for the diffusion coefficient $D_c$ of $Ca^{2+}$ along the grain contact used in the model is $0.01 \cdot D_p$ ($D_p$ is the diffusion coefficient in the pore fluid) at the $T$ and $P$ of the contact fluid. The coefficient of diffusion within the water film, $D_c$, follows an Arrhenius law with an activation energy of 15 kJ/mol (Dysthe et al., 2002b). Considering the uncertainty in $D_c$, we can estimate that the true value probably differs by at least one order of magnitude from the value we have chosen. In Eq. (13) the thickness of the water film at the grain contacts, $\Delta$, is divided by two since it is shared between two contact surfaces. We assume that this water film has a constant thickness of 3 nm (Dysthe et al., 2002b).

The global mass balance equation (Eq. 11) is then coupled to Eq. (14) which represents the local mass balance for each contact on a truncated spherical grain. The concentration of dissolved material within the intergranular fluid phase is given as the difference between the mass produced by dissolution and the mass lost by diffusion:

$$\frac{\partial c_{c,i,Ca}}{\partial t} = \frac{1}{\pi R_{c,i}^2 \frac{\Delta}{2}} \left( \frac{\partial m_{diss,i,Ca}}{\partial t} - \frac{\partial m_{diff,i,Ca}}{\partial t} \right) \qquad (14)$$

In the expression above, $R_{c,i}$ is the radius of the contact surface in the $i$-direction, which is a function of the truncations of the spherical grains. The first term in Eq. (14) represents the local production of dissolved material by dissolution at the grain contacts. The model defines the chemical reaction (i.e. dissolution) of grain-grain contacts in terms of the rate law already presented in Eq. 8, where $R$ is the overall rate (mol m$^{-2}$ s$^{-1}$), $k_{diss}$ is the rate constant, $Q_{c,i}$ is the ion activity product in the contact zone, and $K_{sp,c,i}$ is the calcite solubility product in the intergranular contact zone perpendicular to the $i$ axis ($i=x, y, z$):

$$R = k_{diss}(1 - \Omega) = k_{diss}(1 - Q_{c,i} / K_{sp,c,i}) \qquad (15)$$

Using this law, one can write the flux corresponding to the dissolution process as:

$$\frac{\partial m_{diss,i,Ca}}{\partial t} = k_{diss} \pi R_{c,i}^2 \left( 1 - \frac{c_{c,i,Ca} c_{c,i,CO3}}{K_{sp,c,i}} \right) \qquad (16)$$

In this study, since we simply examine the effect of $p_{CO_2}$ on the rate of PSC, $k_{diss}$ increases linearly by a factor of 3 when $p_{CO_2}$ increases from 0.1 MPa to 2 MPa; for

$p_{CO_2} > 2$ MPa, $k_{diss}$ remains constant (Pokvrovsky et al., 2004). The effect of increasing $p_{CO_2}$ is not limited to an increase in the rate of dissolution (via $k_{diss}$), however, since this also results in a concomitant increase in calcite solubility (i.e. $K_{sp,c}$ and $K_{sp,p}$), which will also affect the rate of PSC.

The grain shape evolution is given by the equations

$$\frac{dL_i}{dt} = -2k_{diss}\bar{V}\left(1 - \frac{c_{c,i,Ca}c_{c,i,CO3}}{K_{sp,c,i}}\right), \quad i = x, y, z \quad (17)$$

$$\frac{dL_f}{dt} = -k_{prec}\bar{V}\left(1 - \frac{c_{p,Ca}c_{p,CO3}}{K_{sp,p}}\right) \quad (18)$$

where $\bar{V}$ is the molar volume of calcite ($3.69 \cdot 10^{-5}$ m³ mol⁻¹) and the calcite precipitation rate $k_{prec}$ is equal to the dissolution rate constant $k_{diss}$. With these equations the code continually updates the texture of the grains and thus couples chemical reactions to grain deformation.

It is important to note that even though the two kinetic processes, dissolution (Eq. 17) and precipitation (Eq. 18), respectively control the grain shape evolution in the contact zones and the pore spaces, these processes are coupled to diffusion in the contact zones, and therefore are not independent. This coupling results in a modification of the chemical affinity terms in both rate equations above (bracketed terms), thereby changing the overall rate of grain shape evolution. As an example, taking the rate of dissolution in Eq. 17, the chemical affinity term $\Omega = \left(1 - \frac{c_{c,i,Ca}c_{c,i,CO3}}{K_{sp,c,i}}\right)$ has the following, respective, limiting minimum and maximum values:

$$0 = \left(1 - \frac{K_{sp,c,i}}{K_{sp,c,i}}\right) < \Omega < \left(1 - \frac{K_{sp,p,i}}{K_{sp,c,i}}\right)$$

- if diffusion is slower than dissolution (i.e. diffusion-limited) $\Omega \to \left(1 - \frac{K_{sp,c,i}}{K_{sp,c,i}}\right) \to 0$

- if dissolution is slower than diffusion (i.e. dissolution-limited), $\Omega \rightarrow \left(1 - \frac{K_{sp,p,i}}{K_{sp,c,i}}\right)$.

The pore space, however, is a special case since dissolved material enters by diffusion from adjacent contact zones, but it can also diffuse to other pore spaces further away.

*2.6. Stress at grain contacts*

The simulation domain is a square rock matrix located at depths of 1, 2 or 3 km and is exposed to constant lithostatic stress and temperature. The normal stress component on an *i*-contact, $\sigma_{c,i}$, is related to the lithostatic stress $\sigma_n$ on the boundaries $\Gamma_i$ by the three relationships which take into account the stress concentration effects due to the porosity of the rock (Dewers and Ortoleva, 1990). These three equations are obtained by circular permutation of the indices in the following relationship:

$$\sigma_{c,i} = \sigma_{n,i} \frac{L_j L_k}{A_i} \tag{19}$$

where *i, j, k* represent the three space directions x, y, and z.

When the lithostatic stress is kept constant, the contact surface area $A_i$ increases as the grains become more and more truncated (Fig. 1). As a consequence the stress normal to this contact decreases with time, and therefore, the driving force for pressure solution also decreases with time. All of these phenomena are due to the continual removal of material at grain-grain contacts by dissolution and the outward diffusion of this dissolved material into the pore fluid.

The final equation updates the porosity, which is defined as the difference between the volume of a cubic box, with lengths $L_x$, $L_y$, $L_z$, and the volume of the truncated grains (Dewers and Ortoleva, 1990). This relationship arises from the textural model of truncated spheres ordered in a centered cubic network (Fig. 1).

*2.7. Numerical methods and boundary conditions*

The equations presented in sections 2.4-2.6 are highly coupled. The deformation of the grains is coupled to transport in the pore fluid through the diffusion step at the contacts and the Fickian diffusion in the pores. Therefore we choose two different numerical modelling techniques. We couple a discrete treatment of the grains (Eqs. 17-

18) with a continuous determination of the dissolved $Ca^{2+}$ in the pore fluid (Eq. 11). The space domain is a 2D rock sample in the *x-z* plane and is modeled as a cubic packing of truncated spheres. It has a thickness of one layer of grains in the *y* direction (Fig. 1). Each element in the numeric grid consists of a chosen number of equal sized grains. The heterogeneity of the rock sample is then given as the variation in grain size between elements. In each element of the grid the number of grains is large (> 100). This allows the treatment of an element as a homogeneous domain with average properties. For simplicity we have assumed constant temperature and no fluid pressure gradient within the simulation domain (i.e. within the pore volume). This assumption is valid because we model processes at the meter-scale and we consider that the hydrodynamics related to the injection step is stabilized. Furthermore, the stress at the boundaries remains constant during the simulations.

Two of the boundaries ($\Gamma_1$ et $\Gamma_4$) of the grid are fixed as stiff walls, where only slip of the simulated domain is allowed. The other two boundaries are free and deform during the simulation. The $Ca^{2+}$ concentration value is fixed at the bottom boundary, and is the same as that in the pore volume at any given time. The boundary conditions for the concentration $c_p$ are imposed on the four boundaries of the system as follows:

$$\frac{\partial c_p}{\partial n} = 0 \text{ on } \Gamma_1, \Gamma_2 \text{ and } \Gamma_3, \tag{20}$$

$$c_p = K_{sp,p} = \text{constant on } \Gamma_4, \tag{21}$$

The geometrical evolution of the grains is modeled as a discrete process where the grains in each grid element are treated separately (see Eqs. 17-18). In order to visualize the deformation of the rock matrix as the grains deform during PSC, an adaptive grid is developed. The nodes change position after each time step as the grains deform. In addition, the contact surface area of each element is updated according to the lengths and radii of the truncated spheres, and is subsequently used to calculate the porosity loss in the *xz*-plane. The total change in volume of each element is then obtained by multiplying the volume variation of each grain by the number of grains in the element. The total deformation of the rock sample is determined by a summation over all the elements.

The mass conservation equation (Eqn. 11) is derived from a continuum model, assuming a continuous solute phase. This equation is solved using a Galerkin finite element method, with linear basis functions on the spatial domain and a Crank-Nicolson scheme in the time domain. All the numerical schemes were programmed in C++ and make use of the numerical library Diffpack (Langtangen, 1999; Gundersen et al., 2002). The program was extensively tested for numerical stability. In all of the simulations, we verified that the total mass of solid was conserved as it should be.

The time step is adjusted such that the maximum deformation is less than 0.1% in all the elements between each time increment. Therefore, the time step automatically adapts in the regions with the most rapid deformation.

## 3. Results

In our model, three different geometries of rock matrix are studied:
- a single homogeneous layer with an initially constant grain size;
- a sedimentary layered rock with an initial local variation of grain size;
- a gouge filling a fracture.

In the results presented here, we aim to compare the evolution of the rock with and without $CO_2$ injection. In order to compare the various simulations, we have chosen a reference case that provides a normalized time scale for the measured PSC rates: the reference (see Figs. 2a and 6) is based on the time needed to achieve a porosity of 5% for a rock at 1 km depth and 40 °C, with an initial grain size of 2 mm ($L_f$ = 1 mm), and a pore fluid that is $CO_2$-free (i.e. $p_{CO_2}$ = $10^{-4.5}$ MPa). In all the simulations, the times scales of deformation are normalized to this reference case (total time for reference deformation equals 740,000 years- see Table 3). In this way we measure the degree to which the sequestration of $CO_2$ perturbs the system and enhances deformation and compaction by PSC. The absolute time scales for PSC deformation for all of the simulations are given in Table 3. Because of uncertainties with respect to several parameters ($k_{diss}$, $k_{prec}$, $D_c$), the absolute time scales related to PSC deformation are difficult to quantify and should be used with caution (see Discussion).

In all cases, both $CO_2$-free and at elevated $p_{CO_2}$, the PSC process is controlled by the diffusion process in the contact zones. This indicates that the intergranular diffusion process is slower than either dissolution in the contact zones or precipitation in the pores. Diffusion in the contact zones, controlled by the $Ca^{2+}$ concentration gradient between the intergranular fluid and the pore fluid, therefore, plays the dominant role in defining the rate of diffusion, the rate of grain indentation, and most importantly, the overall rate of PSC deformation of the rock matrix. This result is predictable, given the rapid kinetics of calcite dissolution and precipitation. Figure 2c, which shows that the $Ca^{2+}$ concentration in the contacts is always greater than that in the pore fluid, graphically demonstrates that the rate of diffusion is the limiting process of PSC for these simulation conditions.

*3.1.    CO$_2$-free simulations*

The rock matrix, located at 1 km depth, consists of grains with a homogeneous size of 2 mm, which corresponds to a coarse oolitic limestone. We consider here a CO$_2$-free pore fluid (i.e. atmospheric $p_{CO_2}$ = 10$^{-4.5}$ MPa). In our model, a typical compaction process without CO$_2$ can last between 10$^4$ to 10$^6$ years, depending on the choice of the parameters $D_c$, $k_{prec}$, $k_{diss}$. As already stated, this configuration represents a reference case for comparison with other simulations at elevated $p_{CO_2}$.

As PSC proceeds, the porosity decreases progressively with time both by grain indentation and by precipitation in the pores (Fig. 2a). As shown in Fig. 2b, $L_f$ increases whereas $L_z$ decreases. With time the grains become truncated and the stress on the contacts decreases. As a consequence, the gradient in [$Ca^{2+}$] between the grain contacts and the pore fluid also decreases (Fig. 2c), leading to a decrease in the rate of overall PSC deformation

For a heterogeneous medium, such as a layered rock, the grain size may vary spatially (Fig. 3). In this case, the PSC rate is faster in the layers with small grains than in layers with coarse grains. Because of this initial grain size difference, the different domains compact at different rates. Layers with smaller grains reach the limit of 5% porosity faster than coarser grain layers. This result is an example of the well-known inverse dependence of pressure solution creep rate on grain size (Rutter, 1976; Gratz 1991). Similar behavior can be observed for a fracture filled with a gouge containing smaller grains (Fig. 4). Both examples show that compaction is homogeneous in the domains with a constant grain size. We do not observe mass transfer between layers as we did for the simulation of PSC in sandstones (Gundersen et al., 2002). This is related to the faster compaction rate in limestones, and the smaller time scales involved for long distance diffusion in the pores.

*3.2.    Effect of elevated partial pressures of $CO_2$*

When $CO_2$ is present at hydrostatic pressure (i.e. $p_{CO_2} = P_p$), PSC is significantly faster compared to the $CO_2$-free case. This situation is clearly visualized in Figure 5 which shows the compaction process for two fractured rocks at 2 km depth, one with $p_{CO_2} = 10^{-4.5}$ MPa (i.e. $CO_2$-free), the other where $p_{CO_2} = P_p$ (see Table 3). The pattern of the porosity reduction is similar in both cases, however the characteristic time scales are very different; at $p_{CO_2} = P_p$, the process is roughly 65 times faster.

We also compare the compaction process at different depths with and without injection of $CO_2$. With no injection of $CO_2$ the rock matrix compacts naturally as PSC progresses. As depth increases, the temperature and the stress increase as well. On the one hand, the effect of stress enhances PSC, mainly by increasing the solubility and the rate of dissolution. On the other hand, increasing the temperature decreases the solubility of calcite (i.e. calcite has retrograde solubility), and this therefore slows down the PSC rate. These two competing factors explain why the PSC rates increase by less than a factor of 3 between 1 and 3 km burial depth (Figure 6a). In a similar manner, for elevated $p_{CO_2}$, the relative increases in the PSC rates with increasing depth are modest (Table 3), again reflecting the competing effects of stress, solubility and kinetics.

When $CO_2$ is present at elevated concentrations, where $p_{CO_2}$ equals the pore fluid pressure, the rates of PSC are greater by factors of ~50-75 (Fig. 6b), as compared to the $CO_2$-free simulations (Fig. 6a), the exact amount depending on depth (Table 3). This translates to a reduction in porosity significantly more rapid than compaction at the same depth in the absence of $CO_2$. This can be explained by two factors. First, at elevated levels of $CO_2$, the pH decreases dramatically (Table 3), leading to increases in calcite solubility by factors of ~50-80, depending on depth. Increasing the values of calcite $K_{sp}$ increases the diffusional gradient between the contact zone and the pores, thereby yielding a higher rate of PSC. Second, elevated $p_{CO_2}$ increases the rates of calcite dissolution (Pokrovsky et al., 2004) and precipitation.

However, an increase in the rate of calcite dissolution in the intergranular zone should not affect the rate of deformation since it appears that the diffusion process is the rate-limiting step in the overall PSC process. On the other hand, an increase in the rate of precipitation in the pores has a positive effect on the rate of diffusion from the contact zone to the pores. What, if any, effect increased levels of $CO_2$ has on the rates of diffusion is unknown.

## 4. Discussion and conclusions

The injection of $CO_2$ into a geological reservoir and the resulting increase in the $p_{CO_2}$ cause acidification of the pore fluids. As the pH of the pore fluids decreases, they become more reactive with respect to the rock matrix; this effect is especially pronounced for carbonates since both the solubility and the reaction kinetics increase dramatically with decreasing pH. This is in accord with the general finding that, to a first approximation, PSC rates depend linearly on the solubility of minerals (Rutter, 1976). Our model shows that for a given depth and temperature, elevated concentrations of dissolved $CO_2$ in the pore fluids lead to rates of PSC deformation that are up to 75 times faster than reference simulations based on $CO_2$-free pore fluids. Thus, our results convincingly demonstrate the direct relation between elevated $p_{CO_2}$ levels, calcite solubility, pore fluid pH, and increased rates of PSC deformation.

Of particular importance to predicting the behavior of future $CO_2$ geological repositories is the accuracy of theoretical PSC deformation rates. Taking just one example from our model, the absolute time scale associated with porosity reduction from 30 to 5% for a carbonate rock matrix with $CO_2$-free pore fluids at 1 km depth (40 °C, 2mm grain size) is on the order of 700,000 years (Table 3). This time frame is most probably unrealistically rapid, given that in nature PSC occurs over much longer time scales. As an example, it is observed that Mesozoic chalks and limestones retain some porosity after several tens of million years, which represents a time period several orders of magnitude greater. There are undoubtedly a myriad number of reasons for this, for which a detailed analysis is beyond the scope of this article. Nonetheless, we briefly discuss how some important parameters influence theoretical PSC deformation rates derived from our model, and how they may contribute to discrepancies with natural PSC rates.

The present study reports 'preliminary results' that are derived from a model that incorporates many simplifications that are both chemical and physical in nature. One of the major simplifications is the lack of coupling between all pertinent chemical species during the deformation process, since at present only $[Ca^{2+}]$ evolves with time, whereas $CaHCO_3^+$, $HCO_3^-$, $CO_3^{2-}$, and $H^+$ remain fixed at their respective initial, equilibrium values. This simplication was necessary in order to reduce numerical noise in the results. While it is difficult to quantify the exact effect of this simplification scheme, it is probable that this results in an overestimation of the PSC rate. One of the main reasons for this is the production of alkalinity via $HCO_3^-$ (see Eqs. 1, 2) during dissolution, resulting in an increase in the pH (i.e. decrease in $[H^+]$) within the contact fluid and pore fluid. The most obvious consequence is a decrease in both calcite solubility and the rate of dissolution with time, which ultimately would decrease the rate of PSC. We are currently developing a code that integrates and fully couples $Ca^{2+}$, $CaHCO_3^+$, $HCO_3^-$, $CO_3^{2-}$, and $H^+$ into a consistent model. Even though a fully coupled model should produce more realistic PSC rates, we estimate that the deformation patterns would not differ too much from those generated in the present study.

The thermodynamic, kinetic, and diffusion parameters that are incorporated in the code play perhaps the most important role in determining the accuracy of theoretical PSC rates of deformation. The rates that we have chosen for the kinetic rate parameters $k_{diss}$ and $k_{prec}$, the value of $n$ (see Eq. 8), and the overall form of the kinetic rate equation(s), determine the rates of dissolution and precipitation in the contact zone and pore space, respectively, and therefore potentially influence the overall rate of PSC deformation. The kinetic parameters we use are not unique and therefore future modeling will include sensitivity analyses based on a variation of these kinetic parameters. In addition, the accuracy of theoretical PSC rates depends on the availability and quality of experimental data, both kinetic and thermodynamic, for conditions relevant to $CO_2$ sequestration. Further refinements in PSC models will depend on additional experimental work to unravel more precisely the effect of elevated $p_{CO_2}$ on $k_{diss}$ and $k_{prec}$ of calcite (and other minerals, as well) at conditions close to equilibrium at elevated $T$ and $P$.

If we consider $CO_2$ sequestration and PSC deformation within the context of conditions in a geological repository, the chemical complexity of natural pore fluids may dictate the need for additional kinetic parameters to be used in the kinetic rate laws. As an example, the presence of certains ions (e.g. $Mg^{2+}$, $SO_4^{2-}$, $PO_4^{3-}$; Svensson and Dreybrodt, 1992; Zuddas and Mucci, 1998; Zhang et al., 2002) or humic acids (e.g. Zuddas et al., 2003) have been shown to inhibit calcite dissolution/precipitation rates. The inhibition of calcite kinetic rates has an important implication for carbonate PSC deformation since the limiting process may switch from diffusion limited to reaction limited. This points out that more experimental work is needed to investigate the effect of fluid chemistry on rates of PSC deformation. The mineralogy of the solid phase is also important. At present, our code only considers a monomineralic (pure calcite) limestone. The effect of impurities, such as clays, potentially has a large effect on PSC deformation. For example, it has been experimentally shown that the presence of clays can dramatically increase rates of PSC (Hickman and Evans, 1995; Renard et al., 2001).

Even though the thermodynamics of the calcite-$H_2O$-$CO_2$ system at STP are well known and quite accurate, future versions of our code need to further develop the temperature and pressure dependencies of most of the equilibria given in Eqs. 3a-h. Accounting for non-ideal behavior at high $p_{CO_2}$ is also necessary with respect to Henry's law constant in Eq. 3b. Moreover, future versions of the code will need to consider equilibria relations in terms of activities and not concentrations; this will be especially important for solutions with elevated ionic strengths. Two other parameters of considerable importance, especially for the case of diffusion-limited PSC deformation, are the rate of diffusion and the thickness (and continuity) of the fluid film in intergranular contacts. Measurement of diffusion coefficients in thin films is experimentally challenging, and therefore the diffusion coefficient and film thickness will remain uncertain quantities that probably contribute significantly to the discrepancy between natural and theoretical PSC deformation rates.

At this stage in the development of our model, many uncertainties exist with respect to the thermodynamic, kinetic, and physical parameters that have been incorporated in the code. Future versions of the code will hopefully be able to address many of these issues. We believe that the main value of the results generated in the present study lies not in the absolute rates, but rather in the relative rates of PSC deformation. Intercomparison of these relative PSC rates will help in better understanding and evaluating the long-term effect of increased rates of PSC on the porosity and viscosity of rock matrices in contact with high $p_{CO_2}$ fluids. Another important benefit of PSC models such as ours is that they may help in focusing laboratory-based PSC experiments. The relation between solubility, pH and PSC rates has important ramifications with respect to future experimental laboratory protocols for investigating the sequestration of $CO_2$. As an example, the injection of subcritical $CO_2$-$H_2O$ mixtures into a rock sample could be simplified by the simple injection of an acidic aqueous fluid, where the pH is adjusted to be equivalent to the pH due to $CO_2$ acidification alone. Nonetheless, such an approach has its limitations and would not be applicable under all circumstances, especially in the case of injection of supercritical $CO_2$ fluids.

**Acknowledgments**

The project has been supported by the French Ministry of Industry through the project PICOR. We would like to thank C. Pequegnat, G. Escorne, and D. Tisserand for their technical assistance.

*Table 1: Parameters used in Eqs. (3-21).*

| Parameter | Description | Units (S.I.) |
| --- | --- | --- |
| $A_i$ $i = x, y, z$ | contact surface area in the $i$-direction | m$^2$ |
| $A_p$ | pore surface area | m$^2$ |
| $c_{c,i,k}$ $i = x, y, z$ | concentration of species $k$ in a grain contact in the $i$-direction | mol/m$^3$ of water |
| $c_{p,k}$ | concentration of species $k$ in the pore fluid | mol/m$^3$ of water |
| $D_c$ | diffusion constant of Ca$^{2+}$ in the grain contact | m$^2$/s |
| $D_p$ | diffusion constant of Ca$^{2+}$ in the pore fluid | m$^2$/s |
| $k_{diss}$ | kinetic rate constant for overall calcite dissolution reaction | mol/(m$^2$s) |
| $k_{prec}$ | kinetic rate constant for calcite precipitation reaction (equal to $k_{diss}$) | mol/(m$^2$s) |
| $k_1, k_2, k_3, k_4$ | kinetic rate constants for individual dissolution reactions | mol/(m$^2$s) |
| $K_{sp,c}$ | calcite solubility product in the grain contacts | (mol/m$^3$)$^2$ |
| $K_{sp,p}$ | calcite solubility product in the pores | (mol/m$^3$)$^2$ |
| $K_1, K_2 \ldots$ | equilibrium constants for the calcite-water system | - |
| $L_i$ $i = x, y, z$ | length of the truncated sphere in the $i$-direction | m |
| $L_f$ | radius of the spherical grain | m |
| $P_c$ | contact zone fluid pressure | Pa |
| $P_p$ | pore fluid pressure | Pa |
| $m_{diff}$ | Ca$^{2+}$ transported by diffusion out of the contact | mol |
| $m_{diss}$ | Ca$^{2+}$ dissolved from the grain contacts | mol |
| $m_{prec}$ | precipitated Ca$^{2+}$ on the pore surface | mol |
| $Q_c$ | ion activity product for calcite dissolution in grain contacts | (mol/m$^3$)$^2$ |
| $Q_p$ | ion activity product for calcite dissolution in pores | (mol/m$^3$)$^2$ |
| $p_{CO_2}$ | partial pressure of carbon dioxide | Pa |
| $R_{c,i}$ $i = x, y, z$ | radius of a grain contact in the $i$-direction | m |
| $t$ | time | s |
| $T$ | temperature | K |
| $\overline{V}$ | molar volume of calcite | m$^3$/mol |
| $\Delta$ | thickness of the water-film in the contact intergranular zone | m |
| $\phi$ | porosity (%) | no units |
| $\sigma_{c,i}$ $i = x, y, z$ | normal stress component in the $i$-direction | Pa |
| $\sigma_{n,i}$ $i = x, y, z$ | normal stress to a grain-surface in the $i$-direction | Pa |

Table 2: Coefficients for calculation of temperature dependence of equilibrium constants in the calcite-$H_2O$-$CO_2$ system. After Langmuir (1997), Table A1.1.

| Reaction | a | b | c | d | e | log K 25°C |
|---|---|---|---|---|---|---|
| $H_2CO_3$ (Eq. 3c) | -356.3094 | -0.06091964 | 21834.37 | 126.8339 | -1684915 | -6.352 |
| $HCO_3^-$ (Eq. 3d) | -107.8871 | -0.03252849 | 5151.79 | 38.92561 | -563713.9 | -10.329 |
| $CO_2$ (Eq. 3b) | 108.3865 | 0.01985076 | -6919.53 | -40.45154 | 669365 | -1.468 |
| $CaHCO_3^+$ (Eq. 3e) | 1209.120 | 0.31294 | -34765.05 | -478.782 | not given | -1.106 |

Table 3: Initial conditions, parameters, and concentrations for the numerical simulations at different depths. Time represents the absolute time required to compact samples from 30% to 5% porosity. Note that the results in Figs. 2-6 are normalized to the $CO_2$-free case of deformation at 1 km depth (i.e. 740,000 years).

| Depth (km) | $\sigma_n$ (MPa) | $P_p$ (MPa) | T (°C) | $p_{CO_2}$ (MPa) | [$Ca^{2+}$] (mol m$^{-3}$) | [$CO_3^{2-}$] (mol m$^{-3}$) | [$HCO_3^-$] (mol m$^{-3}$) | pH | Time (yrs) |
|---|---|---|---|---|---|---|---|---|---|
| 1 | 22 | 10 | 40 | $10^{-4.5}$ | 0.38 | 8.7 e-3 | 0.74 | 8.3 | 740,000 |
| 1 | 22 | 10 | 40 | 10 | 20 | 1.6 e-4 | 57.3 | 4.7 | 13,000 |
| 2 | 44 | 20 | 70 | $10^{-4.5}$ | 0.25 | 8.0 e-3 | 0.49 | 8.3 | 280,000 |
| 2 | 44 | 20 | 70 | 20 | 16.8 | 1.2 e-4 | 47.8 | 4.5 | 4300 |
| 3 | 66 | 30 | 100 | $10^{-4.5}$ | 0.16 | 5.7 e-3 | 0.32 | 8.4 | 190,000 |
| 3 | 66 | 30 | 100 | 30 | 12.6 | 7.4 e-5 | 35.4 | 4.5 | 2500 |

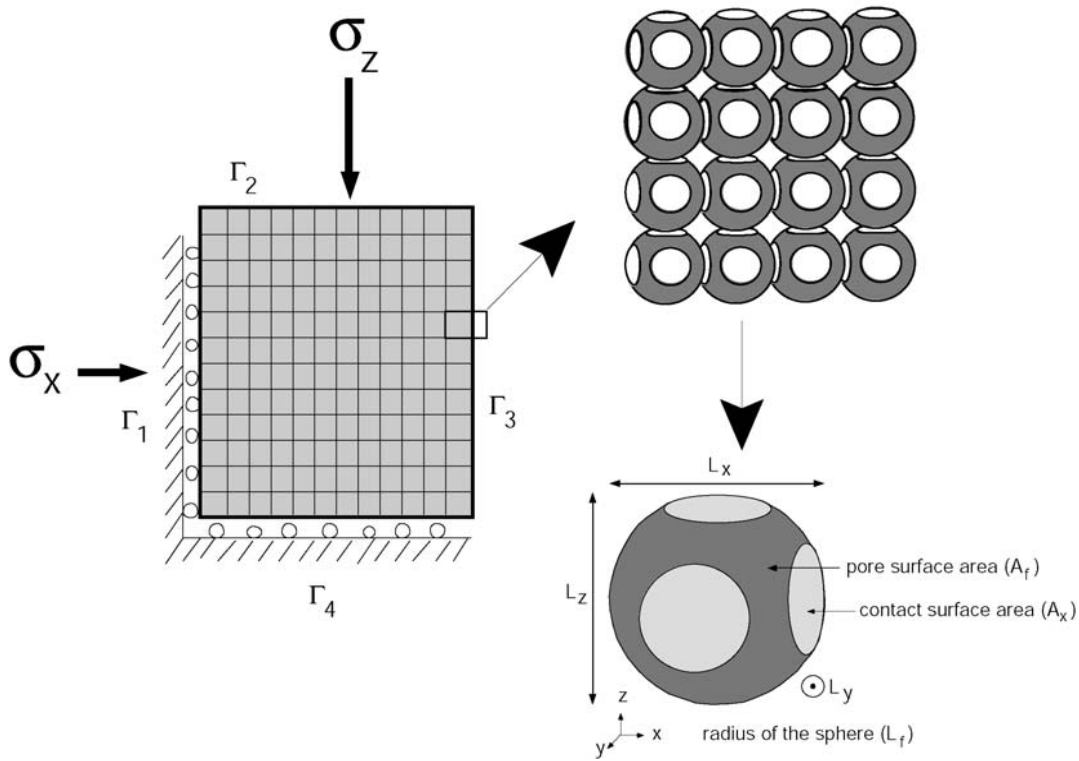

Figure 1: 2D domain where coupled pressure solution creep and local solute transport are simulated. Each element in the numerical grid contains homogeneous grains with a given grain size. The grain size can vary between elements. The domain is 2D (*x* and *z*-directions) and has a layer thickness of one grain in the *y*-direction. In this example the rock consists of layers where each element contains grains of different sizes. The left and bottom boundaries ($\Gamma_1$ and $\Gamma_4$) have no have slip conditions and act as stiff walls, whereas the two other boundaries ($\Gamma_2$ and $\Gamma_3$) are free to deform.

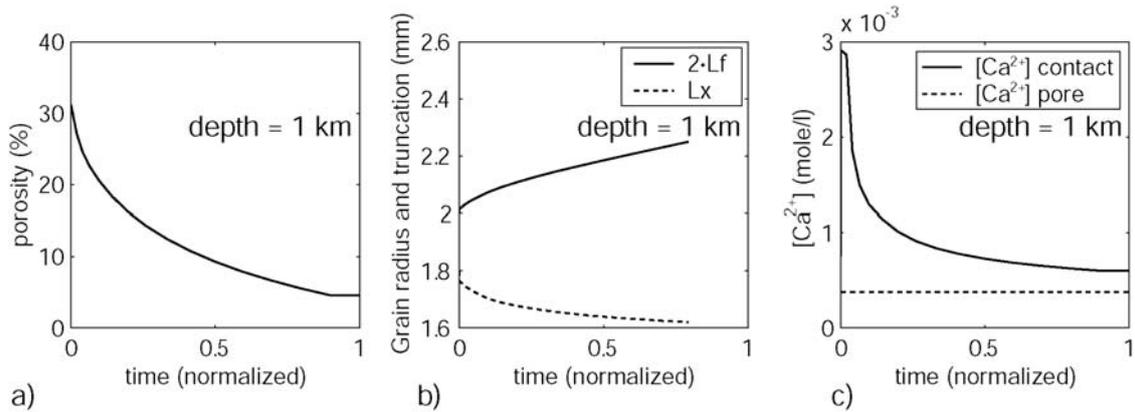

Figure 2: PSC deformation for a limestone located at 1 km depth, 40 °C, with a homogeneous grain size of 2 mm and $CO_2$-free pore fluid (i.e. $p_{CO_2} = 10^{-4.5}$ MPa). a) Porosity reduction with time b) Increase with time of grain radius, $L_f$, as precipitation occurs in the pore space and decrease of grain height, $L_z$, as a result of grain indentation. c) Concentration of $Ca^{2+}$ in the contact and in the pore fluids as a function of time.

The time for the other simulations (Figs. 3-6) is normalized to this simulation, which serves as a reference case.

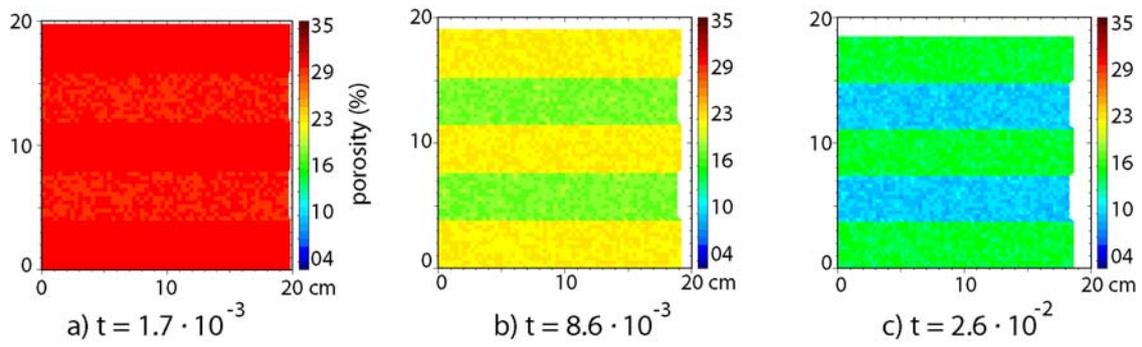

Figure 3: Time evolution of porosity reduction as a result of PSC for a layered matrix at 2 km depth and $p_{CO_2} = 10^{-4.5}$ MPa. The sample was initially 20x20 cm wide. The porosity (in %) decreases faster in layers with initially smaller (1.8 mm ± 0.1 mm; 2 off-centered layers) grains than in layers with coarser grains (2 mm ± 0.1 mm). The time scales in a, b, and c have been normalized with respect to Fig. 2a.

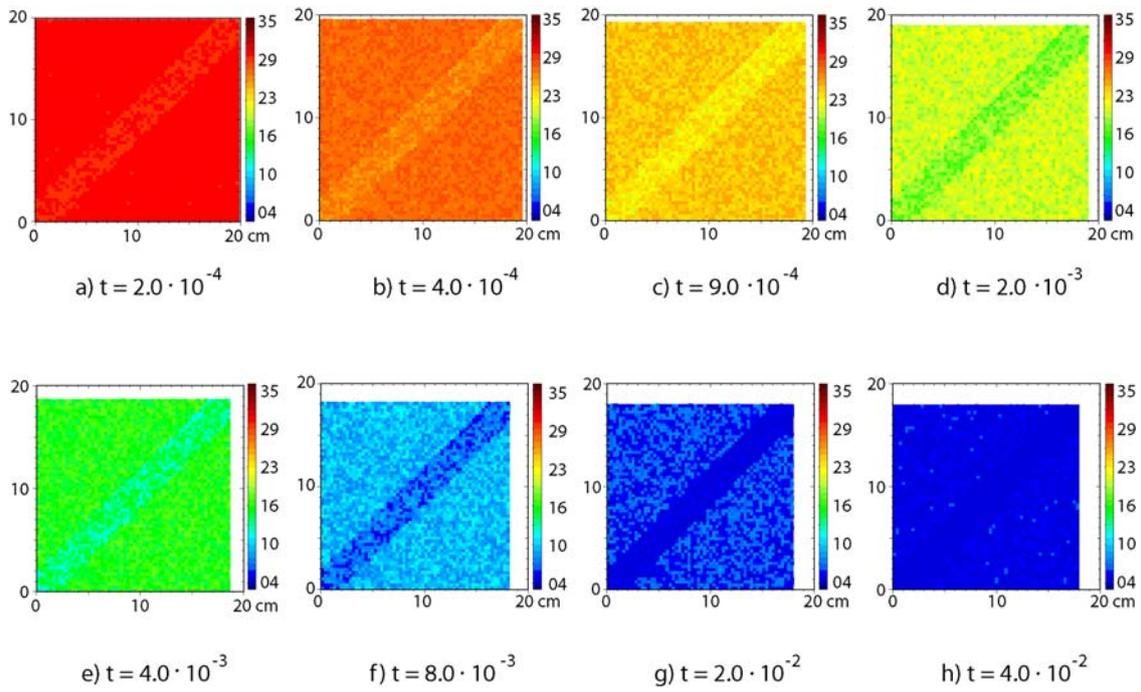

Figure 4: Porosity reduction for a sample containing a fracture. The sample, initially 20x20 cm wide, is located at 2 km depth, $p_{CO_2} = 10^{-4.5}$ MPa. The gouge inside the fracture has a smaller grain size (1 mm ± 0.1 mm instead of 2 mm ± 0.1 mm in bulk) and therefore deforms faster. The normalized time (reference case Fig. 2a) increases from left to right and top to bottom.

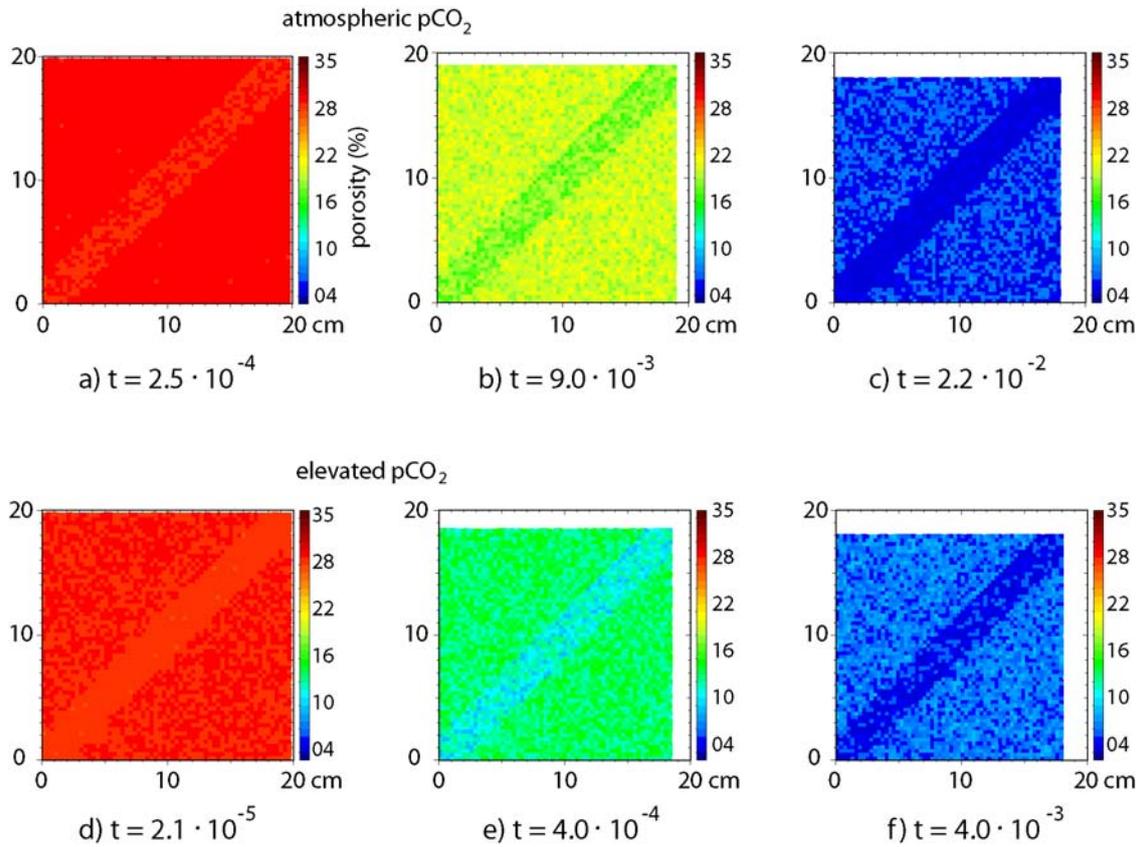

Figure 5: Time evolution of the porosity for a fracture filled with a gouge located at 2 km depth without $CO_2$, $p_{CO_2}$ = $10^{-4.5}$ MPa (top), and with an elevated $CO_2$ concentration, $p_{CO_2}$ = 20 MPa (bottom). The pattern of porosity reduction is similar for the two conditions; however, the required time for porosity reduction is roughly 65 times faster in the presence of $CO_2$. The time is normalized to that in Fig. 2a.

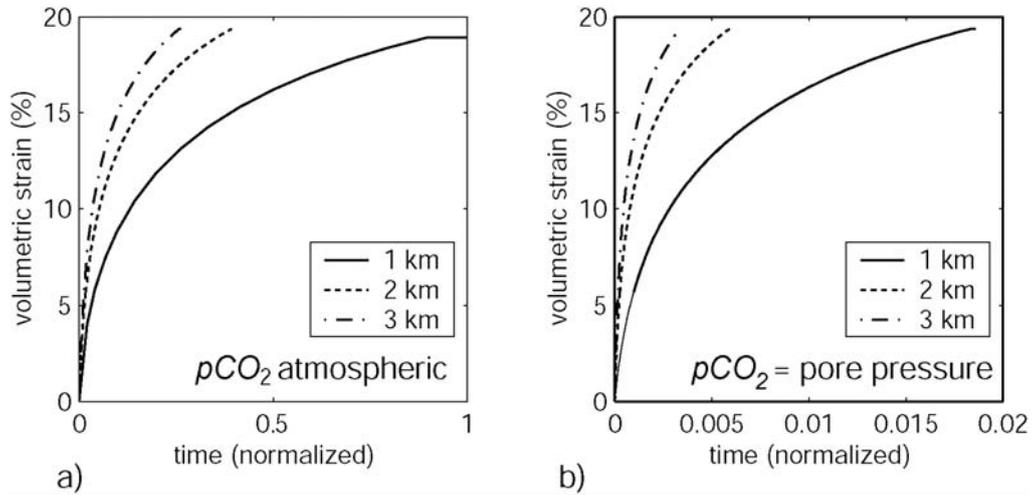

Figure 6: Simulation of the compaction of a homogeneous porous limestone at three different depths, 1, 2 and 3 km, where the geological parameters are given in Table 3. The strain for a 1 km deep system is taken to be the reference for the normalized time. a) volumetric strain, defined as *dV/V*, where *V* is the initial volume of the porous medium, for atmospheric $p_{CO_2}$ ($p_{CO_2} = 10^{-4.5}$ MPa). b) Volumetric strain as a function of time for a system where $p_{CO_2}$ is equal to the hydrostatic pore fluid pressure ($p_{CO_2}$ = 10, 20, 30 MPa). Note that the times required for compaction at a given depth are up to 75 times more rapid than for the corresponding cases at atmospheric $p_{CO_2}$; note, however, that the overall behavior of the compaction curves is similar.